\documentclass[doublecol,a4paper,showpacs]{epl2}
\usepackage{graphicx}
\usepackage[all]{xy}
\usepackage{amsmath}
\usepackage{amssymb}
\usepackage{tensor}
\usepackage[colorlinks=true, pdfstartview=FitV, linkcolor=blue, citecolor=red, urlcolor=black, breaklinks=true]{hyperref}
\newcommand{\be}{\begin{equation}}
\newcommand{\ee}{\end{equation}}
\newcommand{\ben}{\begin{eqnarray}}
\newcommand{\een}{\end{eqnarray}}
\newcommand{\bes}{\begin{subequations}}
\newcommand{\ees}{\end{subequations}}
\def\bal#1\eal{\begin{align}#1\end{align}}

\newcommand{\LL}{{\cal L}}

\begin{document}

\title{New twinlike models for scalar fields}
\author{D. Bazeia\inst{1} \and L. Losano\inst{1} \and M.A. Marques\inst{1} \and R. Menezes\inst{2,1}}
\shortauthor{D. Bazeia \etal}
\institute{                    
  \inst{1} Departamento de F\'\i sica, Universidade Federal da Para\'\i ba, 58051-970 Jo\~ao Pessoa, PB, Brazil\\
  \inst{2} Departamento de Ci\^encias Exatas, Universidade Federal da Para\'{\i}ba, 58297-000 Rio Tinto, PB, Brazil}

\abstract{This work investigates twinlike scalar field models that support kinks with the same energy density and stability. We find the first order equations compatible with the equations of motion. We use them to calculate the conditions under which they attain the twinlike character. The linear stability is also investigated, and there we show that the addition of extra requirements may lead to the same stability under small fluctuations.}
\maketitle

\section{Introduction}
In high energy physics, defect structures appear in scalar field models \cite{wilets,manton,vachaspati} and may engender a topological or nontopological nature. Among these objects, the simplest ones are kinks and lumps, found in $(1,1)$ spacetime dimensions. They appear under the action of a single real scalar field as static solutions of the equations of motion with the presence of nonlinearities that arise from the potential. Usually, kinks (lumps) are stable (unstable) under small fluctuations of the corresponding static field configurations. 

The canonical or standard models that support the aforementioned structures have their Lagrangian densities presenting kinematical and potential terms. Over the more recent years, defect structures have been studied in models with noncanonical features. For instance, in Ref.~\cite{babichev}, Babichev investigated the so called k-defects, which arise in models with modifications in the kinetic term in the Lagrangian density. The inspiration comes from the context of inflation \cite{inf1,inf2}, where nonlinear terms added to the kinetic part of the Lagrangian density may allow for the inflation be driven without the presence of a potential. Since then, many papers appeared in the literature dealing with defect structures in generalized models; see, e.g., Refs.~\cite{genkink,gen1,gen2,gen3,gen4,gen5} and references therein

An interesting feature that may appear in noncanonical models is the twinlike character. In Ref.~\cite{trodden}, the authors introduced a generalized model that has mass parameter and supports the very same solutions and energy density of the standard one; see also Ref.~\cite{altw1}. In this case, it is said that the generalized and standard models are twins. Notwithstanding that, the linear stability of the models differ from each other. So, in Ref.~\cite{tstab1}, the authors introduced a model that also engender the twinlike feature in the stability under specific conditions.

In this work, we introduce a novel model that support similar features from the one in Ref.~\cite{trodden}, which we call ALTW model. But we do more, showing that the present model also support the twinlike character in the stability, with the same eigenfunctions and eigenvalues in the stability equation. We start the investigation by reviewing the basic properties of the standard and ALTW model, such as the equation of motion, first order equation, energy density and linear stability. Next, we introduce a new model and seek for the conditions that lead to the twinlike character. We then end our work by presenting conclusions and perspectives for future research.

\section{Initial Considerations}
Before presenting our new model, let us review the essencial features of the standard case, whose Lagrangian density is
\be\label{ls}
\LL_s = X - U(\phi),\quad\text{where}\quad  X =\frac12\partial_\mu\phi\partial^\mu\phi,
\ee
and $U(\phi)$ denotes the potential. Since our interest is to deal with kinks, we consider static configurations, $\phi=\phi(x)$, in which $X=-{\phi^\prime}^2/2$. The equation of motion that govern the scalar field is
\be
\phi'' = U_\phi,
\ee
where the prime stands for the derivative with respect to $x$ and $U_\phi = dU/d\phi$. The energy density is calculated standardly; it is given by
\be
\rho = \frac12{\phi'}^2 + U.
\ee
One can multiply both sides of the equation of motion by $\phi^\prime$ to show that it can be reduced to the first order, as ${\phi'}^2/2 = U + C$, where $C$ is an integration constant. To ensure the solutions engender finite energy, we take $C=0$ to get
\be\label{fos}
\frac12{\phi'}^2 = U,
\ee
which can be written in the form $X=-U(\phi)$. By using this equation, we can write the energy density as $\rho={\phi'}^2 = 2U(\phi)$. The linear stability of the solutions are investigated through a Schr\"odinger-like equation, which arises from the time dependent equation of motion with the field $\phi(x,t) = \phi(x) + \sum_i\cos(\omega_i t) \eta_i(x)$, where $\eta_i(x)$ denotes the fluctuations around the static solutions $\phi(x)$. The equation that describes the profile of $\eta_i$ is
\be\label{stabs}
\eta_i^{\prime\prime} + U_{\phi\phi}\big|_{\phi=\phi(x)}\eta_i = \omega_i^2\eta_i.
\ee
We call $U_{\phi\phi}\big|_{\phi=\phi(x)}$ the stability potential. In the above equation, the solutions are stable if
$\omega_i\geq0$.

\subsection{The ALTW model}
In Ref.~\cite{trodden}, the authors introduced a model whose solutions and their energy densities are the same of the standard case described by Eq.~\eqref{lagrang}. Its Lagrangian density has the form
\be\label{ltrodden}
\LL = M^2-M^2\sqrt{\left(1+\frac{2U(\phi)}{M^2}\right)\left(1-\frac{2X}{M^2}\right)}\,,
\ee
where $X$ is as in Eq.~\eqref{ls} and $M$ is a mass parameter. The asymptotic behavior of the above Lagrangian density with respect to the mass can be studied by taking a Maclaurin expansion of this expression with respect to $1/M^2$ being very small. By doing this, we get
\be
\LL = X-U(\phi) +\frac{1}{2M^2}\left(X+U(\phi)\right)^2 + \mathcal{O}\left(\frac{1}{M^4}\right).
\ee
So, the generalized model in Eq.~\eqref{ltrodden} approaches to the standard scenario in Eq.~\eqref{ls} as $M$ gets larger and larger. In this sense, we call $U(\phi)$ the potential of the modified model. To work with this generalized model, we follow Ref.~\cite{genkink}. For static configurations, one can show the equation of motion is given by
\be
\left(\sqrt{\frac{M^2+2U}{M^2-2X}}\,\phi^\prime\right)^\prime = \sqrt{\frac{M^2-2X}{M^2+2U}}\,U_\phi.
\ee
Regarding the energy density, we have
\be
\rho = M^2\sqrt{\left(1+\frac{2U(\phi)}{M^2}\right)\left(1+\frac{{\phi^\prime}^2}{M^2}\right)}-M^2.
\ee
The key feature of this model is that it supports the very same first order equation of the standard model, which can be seen in Eq.~\eqref{fos}. Thus, it engender the same defect structures. Moreover, since Eq.~\eqref{fos} is valid here, the above energy density can be rewritten as $\rho = 2U(\phi)$, which is exactly the same of the standard case. Since the models share the same solutions and energy density, they are called twins. One may investigate the linear stability similarly as in the standard case to find that we get a Sturm-Liouville eigenvalue equation in this case, written as
\be
\begin{aligned}
	&-\left(M^2\sqrt{\frac{M^2+2U}{(M^2-2X)^3}}\,\eta_i^\prime\right)^\prime=\Bigg(\sqrt{\frac{M^2-2X}{(M^2+2U)^3}}\,U_\phi^2\\
	&-\sqrt{\frac{M^2-2X}{M^2+2U}}\,U_{\phi\phi} + \left(\frac{U_\phi \phi^\prime}{\sqrt{(M^2+2U(\phi))(M^2-2X)}}\right)^\prime
	\\
	&+\sqrt{\frac{M^2+2U}{M^2-2X}}\,\omega^2\Bigg)\eta_i.
\end{aligned}
\ee
So, even considering the first order equation \eqref{fos}, the stability of the solution in the modified model is not the same of the standard case, since this equation differs from Eq.~\eqref{stabs}. The model described by the Lagrangian density \eqref{ltrodden} was generalized in Ref.~\cite{troddengen}, where it was found a class of twinlike models that presents a mass parameter. However, their linear stabilities are not the same.

Twinlike models that engender the same linear stability were investigated in Refs.~\cite{tstab1,tstab2,tstab3}. In Ref.~\cite{twinstab4}, we have studied the conditions to obtain twinlike models for kinks, vortices and monopoles with the same stability up to an arbitrary order. In these models, however, there is no mass parameter as in Lagrangian density \eqref{ltrodden} to connect the generalized model to the standard one.

\section{New Model}
We now introduce a novel Lagrangian density that engender the same kinklike solutions, energy density and stability of the standard case with the presence of a mass parameter. The model is similar to the ALTW model \cite{trodden} defined in \eqref{ltrodden} and is motivated by the previous work \cite{tstab1}, where the twinlike model is extended to work with the same stability behavior, and also by \cite{twinstab4}, in which one investigates several twinlike possibilities with kinks, vortices and monopoles. It is given by 
\be\begin{aligned}\label{lagrang}
\LL &= -\frac{M^2+2\left(U+R(U)\right)}{2}F\left(Y\right) \\
&\hspace{4mm}+a_1\, X +a_2\, Q(X)+b_1\, U+b_2\, R(U)+C,
\end{aligned}
\ee
with $a_1$, $a_2$, $b_1$, $b_2$ and $C$ being real parameters, $X$ as in Eq.~\eqref{ls} and 
\be
Y =\frac{M^2-2(X-Q(X))}{M^2+2(U+R(U))}.
\ee
The Lagrangian density \eqref{lagrang} is a generalization of a class of twinlike models introduced in Ref.~\cite{tstab1}. Here, we unveil a novel possibility, in which the quantity $Y$ presents the functions $Q(X)$ and $R(U)$. The standard case in Eq.~\eqref{ls} may be recovered by taking $Q(X)=0$, $R(U)=0$, $F(Y)=\alpha-\beta+\beta\,Y$, $a_1=1-\beta$ and $b_1=\alpha-\beta-1$. Moreover, we also include the parameter $M$ similarly as it was done in Ref.~\cite{trodden} with the model in Eq.~\eqref{ltrodden}.

In Ref.~\cite{tstab1}, the generalized model admits the same solutions of the standard case, which are given by $X=-U$, for $Y=1$, matching with Eq.~\eqref{fos}. Since our model contains additional functions, $Q(X)$ and $R(U)$, we impose that, for $Y=1$, 
\be\label{QRconst}
R(U)=Q(X=-U).
\ee
This gives many possibilities for the functions $Q(X)$ and $R(U)$. For instance, we can take the pairs $Q(X)= -\sin (X)$ and $R(U)=\sin(U)$, or $Q(X)= \cos (X)$ and $R(U)=\cos(U)$. We also take the values associated to the function $F$ and its derivatives as
\be\label{cond} 
F(1)=\alpha,\quad F_Y(1)=\beta \quad \text{and} \quad F_{YY}(1)=\gamma,
\ee
where $\alpha$, $\beta$ and $\gamma$ are real.

Similarly to the model in Ref.~\cite{trodden} described by Eq.~\eqref{ltrodden}, the Lagrangian density \eqref{lagrang} can be seen as a generalization of the standard case in the sense that, for $M$ very large, it tends to behave as
\be
\LL_{asy}=X-U-\frac{\gamma}{M^2} \Big(X-Q(X)+U+R(U)\Big)^2 + \mathcal{O}\left(\frac{1}{M^4}\right).
\ee
Thus, the standard case in Eq.~\eqref{ls} is obtained through the limit $1/M\to0$ in Eq.~\eqref{lagrang}.

The equation of motion associated to the Lagrangian density in Eq.~\eqref{lagrang} is
\be\label{eomkink}
\begin{aligned}
&\partial_\mu\big((a_1+F_Y+(a_2-F_Y)Q(X))\partial^\mu\phi\big)\\
&+ \left(F-b_1-YF_Y\right)U_\phi+ \left(F-b_2-YF_Y\right)R_\phi=0.	
\end{aligned}
\ee
We also may take advantage of the invariance with respect to spacetime translations to calculate the energy momentum tensor, which is written as
\be
T_{\mu\nu} = \left(a_1+F_Y+(a_2-F_Y)Q(X)\right) \partial_\mu\phi\partial_\nu\phi - \eta_{\mu\nu}\LL.
\ee

To investigate the presence of defect structures, we proceed as before and consider static configurations, $\phi=\phi(x)$, which, again, leads to $X=-{\phi^\prime}^2/2$. In this case, the equation of motion \eqref{eomkink} reads:
\be\label{eomstatickink}
\begin{aligned}
	&\left(\left(a_1+F_Y+(a_2-F_Y)Q(X)\right)\phi^\prime\right)^\prime\\
& = \left(F-b_1-YF_Y\right)U_\phi+\left(F-b_2-YF_Y\right)R_\phi.
\end{aligned}
\ee
For static solutions, the non null components of the energy-momentum tensor are the energy density, $\rho=T_{00}$, and the stress $\sigma=T_{11}$. They are respectively given by
\bes
\bal\label{edens}
\rho(x) &= \frac12(M^2+2(U+R(U)))F(Y)-a_1X-a_2Q(X)\nonumber\\
		&\hspace{4.1mm}-b_1U-b_2R(U)-C,\\
\sigma(x) &= -\frac{B}2 F(Y)+a_1X+a_2Q(X)+b_1U+b_2R(U)\nonumber\\
		&\hspace{4.1mm}+C-2X\left(a_1+F_Y+(a_2-F_Y)Q(X)\right).
\eal
\ees
The stability under contractions and dilations requires the stressless condition; see Ref.~\cite{genkink}. By setting $\sigma=0$, we get a first order equation
\be\label{foalg}
\begin{aligned}
	&-\frac12 (M^2+2(U+R(U)))F(Y)+a_1X+F_Y Q(X)\\
&+b_1 U+b_2R(U)+C-2X(F_Y+a_1)=0.
\end{aligned}
\ee
Since $X=-{\phi^\prime}^2/2$, this is a first order differential equation. One can take the derivative of the above expression to show that this first order equation solves the equation of motion \eqref{eomstatickink}. On the other hand, this can be seen as an algebraic equation that relates the functions $X$ and $U$. We remark here that the latter equation, which comes from the stressless condition, $\sigma=0$ (see Ref.~\cite{genkink}), is very important in the construction of the model described by the Lagrangian density in Eq.~\eqref{lagrang}. It only presents $X$ and $U$, that are the quantities that one uses to get the same solutions of Eq.~\eqref{fos}, which is also represented by $X=-U$. This is a feature that must be present in order to obtain twinlike models. Since the conditions \eqref{cond} are satisfied, one can show that $X=-U$ is a solution for
\be\label{slp}
C=\frac12 \alpha M^2,\quad b_1=\alpha-2\beta-a_1,\quad b_2=\alpha-\beta.
\ee
Under these conditions, the first order equation \eqref{foalg} becomes the first order equation \eqref{fos}. Also, it is ease to verify that the Lagrangian density \eqref{lagrang} support the same solutions of the standard case, described by Eq.~\eqref{ls}.

We now focus on the energy density. By imposing the conditions \eqref{slp} in Eq.~\eqref{edens}, we get
\be
\rho(x) = 2(\beta+a_1)U+(\beta-a_2)R(U).
\ee
We then impose the conditions $a_1=1-\beta$ and $a_2=\beta$ to get the same energy density of the standard case, i.e., $\rho=2U$. Since the models present the same solutions and energy density, we say that the family of models described by \eqref{lagrang} with \eqref{slp} and the standard case are twins. The conditions to obtain twinlike models are summarized as $C=\alpha M^2/2$ and
\be\label{condtwin}
a_1=1-\beta,\quad a_2=\beta,\quad b_1=\alpha-\beta-1,\quad b_2=\alpha-\beta,
\ee
where the parameters $\alpha$ and $\beta$ come from Eq.~\eqref{cond} and the functions $Q(X)$ and $R(U)$ must obey the constraint in Eq.~\eqref{QRconst}.

\subsection{Linear Stability}
Before ending the investigation, let us focus on the linear stability of the solutions. To do so, we consider
time dependent small fluctuations around the static solutions, writing the field in the form
$\phi(x, t) = \phi(x) + \sum_i \cos(\omega_i t)\eta_i(x)$, 
where $\phi(x)$ is the solution of the static equation \eqref{eomstatickink}.
Replacing this in the time dependent equation of motion \eqref{eomkink} and considering terms up to first order in $\eta_i(x)$, a lengthy but straightforward calculation lead us to the stability equation
\be\label{SL}
	\left(a(x)\,\eta_i^\prime\right)^\prime + b(x)\,\eta_i= c(x)\,\omega_i^2 \eta_i,
\ee
where
\bes
\bal
a(x) &= a_1+F_Y+\big(a_2-F_Y\big)Q_X\nonumber\\
     &+\phi^{\prime\,2}\Big(\frac{2}{B}\big(1-Q_X\big)^2-\big(a_2-F_Y\big)Q_{XX}\Big), \\
b(x) &= \big(F-b_1-YF_Y\big)U_{\phi\phi}-\big(F-b_2-YF_Y\big)R_{\phi\phi}\nonumber\\
	 &+\bigg(\frac{2}{B}\big(1-Q_X\big)\big(U_{\phi}+R_{\phi}\big)\phi^{\prime}YF_{YY}\bigg)^{\prime}\nonumber\\
	 &+\frac{2}{B}\big(U_{\phi}+R_{\phi}\big)^2Y^2F_{YY},\\
c(x) &= a_1+F_Y+\big(a_2-F_Y\big)Q(X),
\eal\ees
and
\be 
B =M^2+2(U+R(U)).
\ee
We remark here that, as we have shown in Ref.~\cite{sl}, one can write the above Sturm-Liouville equation in terms of supersymmetric partners which, for kink solutions, are finite and regular. This means that the above equation only admits non negative eigenvalues, so kinks are stable under small fluctuations. To get a better understanding of the stability, let us follow the route proposed in Ref.~\cite{genkink} to show that the above equation may be transformed into a Schr\"odinger-like one. First, we define the quantity
\be\label{st1}
A^2= 1+ \frac{2(1-Q_X)^2F_{YY}\phi^{\prime\,2}+(a_2-F_Y)B\phi^{\prime\,2}Q_{XX}}{B\left(a_1+F_Y+(a_2-F_Y)Q_X\right)},
\ee
which is associated to the hyperbolicity of the stability equation. By performing the change of variables 
\bes
\bal\label{st2} 
& dx=A\,dz\\
&\eta_i=\frac{u_i}{\sqrt{\big(a_1+F_Y+\big(a_2-F_Y\big)Q_X\big)A}},
\eal
\ees
one can show the Sturm-Liouville equation \eqref{SL} is transformed into a Schr\"odinger-like one, in the form
\be
-u_{i,zz}+U(z) u_i= \omega^2 u_i,
\ee
in which the stability potential is given by
\be\label{st3}
\begin{aligned}
	U(z) &= \frac{\left(\sqrt{\left(a_1+F_Y+(a_2-F_Y)Q_X\right) A}\right)_{zz}}{\left(\sqrt{\left(a_1+F_Y+(a_2-F_Y)Q_X\right) A}\right)}\\
&+ \frac{1}{a_1+F_Y+(a_2-F_Y)Q_X} \Bigg(\frac{2(U_{\phi}+R_{\phi})^2 Y^2}{B}\\
&\times F_{YY}+(F-b_1-YF_Y)U_{\phi\phi}+(F-b_2-YF_Y)\\
& \times R_{\phi\phi}+\frac1{A}\bigg(\frac{2(1-Q_X)(U_{\phi}+R_{\phi})YF_{YY}}{B}\frac{\phi_z}{A}\bigg)_z\Bigg),\\
\end{aligned}
\ee
where the subscript $z$ represents derivatives with respect the variable $z$. We now impose $Y=1$ and the conditions in Eqs.~\eqref{cond} and \eqref{condtwin}. By doing so, the above stability potential simplifies to
\be\label{st3}
\begin{aligned}
	U(z)& = \frac{\left(\sqrt{A}\right)_{zz}}{\left(\sqrt{A}\right)}+\frac{2\gamma(U_{\phi}+R_{\phi})^2}{B}+U_{\phi\phi}\\
&+\frac1{A}\bigg(\frac{2\gamma(1-Q_X)(U_{\phi}+R_{\phi})}{B}\frac{\phi_z}{A}\bigg)_z,\\
\end{aligned}
\ee
with $A^2$ in Eq.~\eqref{st1} becoming
\be
A^2= 1+ \frac{2\gamma(1-Q_X)^2\phi^{\prime\,2}}{B}.
\ee
We then see that the conditions \eqref{QRconst}, \eqref{cond} and \eqref{condtwin} are not enough to make the model \eqref{lagrang} support the same stability of the standard one in Eq.~\eqref{ls}. This occurs due to the presence of the parameter $\gamma$, associated to the condition $F_{YY}(1)=\gamma$. So, for $\gamma\neq0$, we obtain models that engender the same solution and energy density, but distinct stability equations. This is what occurs in Refs.~\cite{trodden,troddengen}. Here, the presence of the function $F(Y)$ allows us to take $\gamma=0$, i.e., $F_{YY}(1)=0$ to obtain models with the same solution, energy density and linear stability. So, in order to distinguish the models, one must consider higher order stabilities. We remark here that one may follow the lines of Ref.~\cite{twinstab4} and try to find conditions that generalizes the twinlike character up to a given stability order, but this is out of the scope of the present work. 

\section{Conclusion}\label{sec5}
In this paper, we have studied a generalized scalar field model that presents the very same solution, energy density and stability of the standard model. We have reviewed the basic properties of the standard model, including the presence of a first order equation to describe the field profile; in this case, the linear stability is driven by an eigenvalue equation of the Schr\"odinger type. Next, we have also looked at the basic features of the ALTW model proposed in Ref.~\cite{trodden}, which engender a mass parameter that allows for the approximation of this model with the standard case for very large values of the mass. The solution and energy density of interest are the very same of the standard case, so the models are twins. Notwithstanding that, the linear stability is investigated with a Sturm-Liouville equation, and the model does not support the same linear stability of the standard case.

In order to extend the twinlike character to the stability, we have introduced the model \eqref{lagrang}, which is an extension of the models investigated in Refs.~\cite{trodden,troddengen,tstab1}. It presents a function $F(Y)$, such that, differently from the previous works, $Y$ presents general functions that are not obligated to be linear. This opened up a myriad of possibilities; see below Eq.~\eqref{QRconst}. Similarly to the ALTW model, one can also obtain the standard model by taking the limit of infinite mass. We then have worked the model out to show that, for the specific conditions in Eqs.~\eqref{QRconst}, \eqref{cond} and \eqref{condtwin}, the generalized model \eqref{lagrang} and the standard model \eqref{ls} are twins. We have also studied the linear stability, which is driven by a Sturm-Liouville equation. We were able to transform it into a Schr\"odinger-like one with a change of variables. By imposing an additional condition for $F_{YY}(1)$, we were able to attain the twinlike feature in the stability.

We hope this study fosters other investigations in the area. We remark that, even though we have obtained a generalized model with the same solution, energy density and stability of the standard model, there are some features to be investigated. For instance, one may follow the lines of Refs.~\cite{f1,f2} and calculate the force between defect structures separated by a given distance to verify if the twinlike character is preserved for this quantity. As another perspective, one may also study the scattering of the defect structures in the generalized model \cite{s1,s2,s3,s4}, to see how the scattering changes compared to the standard model. Moreover, one may try to extend this investigation for vortices and monopoles and, in the curved spacetime, for the braneworld scenario with a single extra dimension of infinite extent \cite{b1,b2,b3}. Another issue concerns the presence of models with twinlike behavior in Cosmology, as considered in Refs. \cite{bazeia,adam,dantas}. In \cite{adam} in particular, the authors considered twinlike models to extend some analytical tools like the slow-roll expansion to the case of generalized models. In Ref.~\cite{dantas}, the authors studied canonical and tachyonic models, in the context of dark energy, that support the same acceleration parameter, energy density and pressure.  These and other related issues are currently under consideration, and we hope to report on them in the near future.

\acknowledgements{The work is supported by the Brazilian agencies Coordena\c{c}\~ao de Aperfei\c{c}oamento de Pessoal de N\'ivel Superior (CAPES), grant No.~88887.463746/2019-00 (MAM),  Conselho Nacional de Desenvolvimento Cient\'ifico e Tecnol\'ogico (CNPq), grants Nos. 303469/2019-6 (DB), 404913/2018-0 (DB), 303824/2017-4 (LL) and 306504/2018-9 (RM), and by Paraiba State Research Foundation (FAPESQ-PB) grants Nos. 0003/2019 (RM) and 0015/2019 (DB).}

\end{document}